\documentclass[%
 reprint,
superscriptaddress,
 amsmath,amssymb,
 aps,
]{revtex4-2}

\usepackage{graphicx}
\usepackage{dcolumn}
\usepackage{bm}
\RequirePackage{graphicx}
\RequirePackage{subfigure}
\usepackage{rotating}
\usepackage{amsmath}
\usepackage{mathtools}
\RequirePackage[colorlinks,citecolor=blue,urlcolor=blue,linkcolor=blue]{hyperref}
\usepackage[]{units}
\usepackage{xspace}
\usepackage[switch]{lineno}
\usepackage{comment}

\newcommand{\DBD}{0$\nu\beta\beta$}

\usepackage[switch]{lineno} 
\usepackage{lipsum}

\begin{document}

\preprint{APS/123-QED}

\title{Search for Majoron-like particles with CUPID-0}
\newcommand{\sapienza}{\affiliation{Dipartimento di Fisica, Sapienza Universit\`a di Roma, P.le Aldo Moro 2, 00185, Roma, Italy}}
\newcommand{\infnroma}{\affiliation{INFN, Sezione di Roma, P.le Aldo Moro 2, 00185, Roma, Italy}}
\newcommand{\cnr}{\affiliation{Consiglio Nazionale delle Ricerche, Istituto di Nanotecnologia, c/o Dip. Fisica, Sapienza Università di Roma, 00185, Rome, Italy}}
\newcommand{\lnl}{\affiliation{INFN  Laboratori Nazionali di Legnaro, I-35020 Legnaro (Pd) - Italy}}
\newcommand{\lngs}{\affiliation{INFN  Laboratori Nazionali del Gran Sasso, I-67100 Assergi (AQ) - Italy}}
\newcommand{\lbl}{\affiliation{Lawrence Berkeley National Laboratory , Berkeley, California 94720, USA}}
\newcommand{\infnge}{\affiliation{INFN  Sezione di Genova, I-16146 Genova - Italy}}
\newcommand{\unige}{\affiliation{Dipartimento di Fisica, Universit\`{a} di Genova, I-16146 Genova - Italy}}
\newcommand{\infnmib}{\affiliation{INFN  Sezione di Milano - Bicocca, I-20126 Milano - Italy}}
\newcommand{\unimib}{\affiliation{Dipartimento di Fisica, Universit\`{a} di Milano - Bicocca, I-20126 Milano - Italy}}
\newcommand{\csnsm}{\affiliation{CNRS/CSNSM, Centre de Sciences Nucl$\acute{e}$aires et de Sciences de la Mati$\grave{e}$re, 91405 Orsay, France}}
\newcommand{\cea}{\affiliation{IRFU, CEA, Universit$\acute{e}$ Paris-Saclay, F-91191 Gif-sur-Yvette, France}}
\newcommand{\gssi}{\affiliation{Gran Sasso Science Institute, 67100, L'Aquila - Italy}}
\newcommand{\usc}{\affiliation{Department of Physics  and Astronomy, University of South Carolina, Columbia, SC 29208 - USA}}
\newcommand{\mpi}{\affiliation{Max-Planck-Institut für Physik, D-80805 München, Germany}}
\newcommand{\dis}{\affiliation{DISAT, Universit\`a dell'Insubria, 22100 Como, Italy}}
\newcommand{\JYU}{\affiliation{University of Jyv\"askyl\"a, Department of Physics, P. O. Box 35 (YFL), FI-40014, Finland}}
\newcommand{\FIER}{\affiliation{Finnish Institute for Educational Research, P.O.Box 35 FI-40014 University of  Jyv\"askyl\"a - Finland}}
\newcommand{\CTP}{\affiliation{Center for Theoretical Physics, Sloane Physics Laboratory, Yale University, New Haven, Connecticut 06520-8120 - USA}}
\newcommand{\QUEEN}{\affiliation{Department of Physics and Engineering Physics Astronomy, Queen’s University Kingston, Ontario, K7L 3N6 Kingston, Canada}}

\author{O.~Azzolini}\lnl
\author{J.W.~Beeman}\lbl
\author{F.~Bellini}\sapienza\infnroma
\author{M.~Beretta}\altaffiliation{Present address: Department of Physics, University of California, Berkeley, CA 94720, USA}\unimib\infnmib
\author{M.~Biassoni}\infnmib
\author{C.~Brofferio}\unimib\infnmib
\author{C.~Bucci} \lngs
\author{S.~Capelli}\unimib\infnmib
\author{V.~Caracciolo}\altaffiliation{Present address: Dipartimento di Fisica, Universit\`{a} di Roma Tor Vergata, I-00133, Rome, Italy } \lngs
\author{L.~Cardani}\infnroma
\author{P.~Carniti}\unimib\infnmib
\author{N.~Casali}\infnroma
\author{E.~Celi}\gssi\lngs
\author{D.~Chiesa}\unimib\infnmib
\author{M.~Clemenza}\unimib\infnmib
\author{I.~Colantoni}\infnroma\cnr
\author{O.~Cremonesi}\infnmib
\author{A.~Cruciani}\infnroma
\author{A.~D'Addabbo} \lngs
\author{I.~Dafinei}\infnroma
\author{S.~Di~Domizio}\unige\infnge
\author{V.~Dompè}\sapienza\infnroma
\author{G.~Fantini}\sapienza\infnroma
\author{F.~Ferroni}\infnroma\gssi
\author{L.~Gironi}\unimib\infnmib
\author{A.~Giuliani}\csnsm
\author{P.~Gorla} \lngs
\author{C.~Gotti}\infnmib
\author{G.~Keppel}\lnl
\author{J.~Kotila}\JYU\FIER\CTP
\author{M.~Martinez}\altaffiliation{Present address: Centro de Astropart\'iculas y Física de Altas Energ\'ias, Universidad de Zaragoza, and ARAID, Fundaci\'on Agencia Aragonesa para la Investigaci\'on y el Desarrollo, Gobierno de Arag\'on, Zaragoza 50018, Spain}\sapienza\infnroma
\author{S.~Nagorny}\altaffiliation{Present address: Department of Physics $\&$ Engineering Physics Astronomy, Queen's University Kingston, Ontario, K7L 3N6 Kingston, Canada}\lngs
\author{M.~Nastasi}\unimib\infnmib
\author{S.~Nisi}\lngs
\author{C.~Nones}\cea
\author{D.~Orlandi}\lngs
\author{L.~Pagnanini}\gssi\QUEEN\lngs
\author{M.~Pallavicini}\unige\infnge
\author{L.~Pattavina}\altaffiliation{Present address: Physik-Department and Excellence Cluster Origins, Technische Universit{\"a}t M{\"u}nchen, 85747
Garching, Germany}\lngs
\author{M.~Pavan}\unimib\infnmib
\author{G.~Pessina}\infnmib
\author{V.~Pettinacci}\infnroma
\author{S.~Pirro}\lngs
\author{S.~Pozzi}\unimib\infnmib
\author{E.~Previtali}\unimib\lngs
\author{A.~Puiu}\lngs\gssi
\author{A.~Ressa}\email[Corresponding author: ]{alberto.ressa@roma1.infn.it}\sapienza\infnroma
\author{C.~Rusconi}\lngs\usc
\author{K.~Sch\"affner}\altaffiliation{Present address: Max-Planck-Institut f{\"u}r Physik, 80805 M{\"u}nchen - Germany}\lngs
\author{C.~Tomei}\infnroma
\author{M.~Vignati}\sapienza\infnroma
\author{A.~S.~Zolotarova}\cea

\collaboration{CUPID-0 Collaboration}

\date{\today}

\begin{abstract}
We present the first search for the Majoron-emitting modes of the neutrinoless double $\beta$ decay (\DBD$\chi_0$) using scintillating cryogenic calorimeters. We analysed the CUPID-0 Phase I data using a Bayesian approach to reconstruct the background sources activities, and evaluate the potential contribution of the  $^{82}$Se \DBD$\chi_0$. We considered several possible theoretical models which predict the existence of a Majoron-like boson coupling to the neutrino. The energy spectra arising from the emission of such bosons in the neutrinoless double $\beta$ decay have spectral indices $n=$ 1, 2, 3 or 7. We found no evidence of any of these decay modes, setting a lower limit (90\% of credibility interval) on the half-life of 1.2 $\times$ 10$^{23}$ yr in the case of $n=$ 1, 3.8 $\times$ 10$^{22}$ yr for $n=$ 2, 1.4 $\times$ 10$^{22}$ yr for $n=$ 3 and 2.2 $\times$ 10$^{21}$ yr for $n=$ 7. These are the best limits on the \DBD$\chi_0$ half-life of the $^{82}$Se, and demonstrate the potentiality of the CUPID-0 technology in this field.

\end{abstract}

\maketitle

\section{Introduction}
\begin{figure}[htbp]
\centering
\includegraphics[scale=0.45]{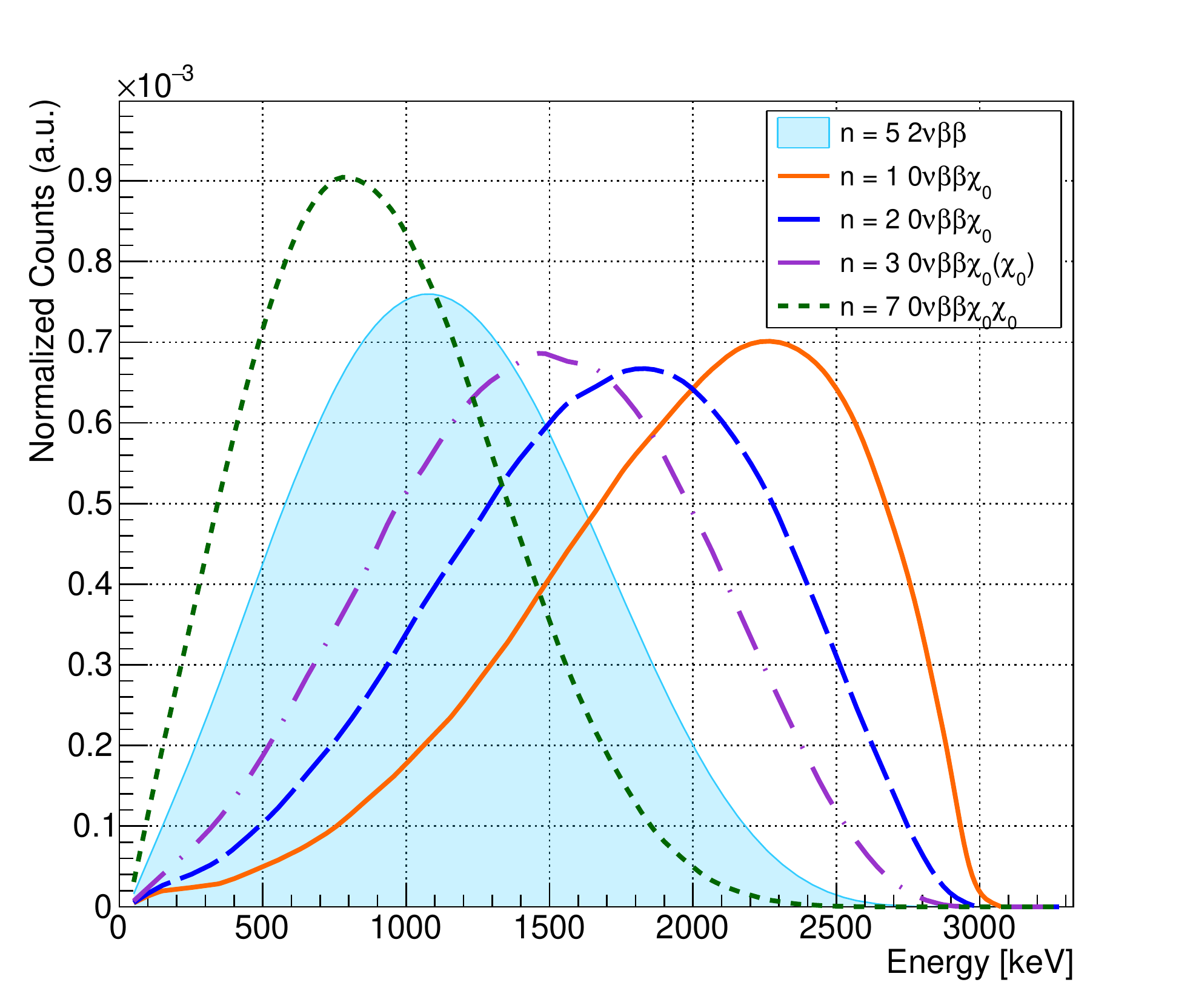}
\caption{Predicted energy spectrum of the $^{82}$Se \DBD$\chi_0$ decay. The abscissa refers to the sum of the energy of the electrons emitted in the decay. We show all the Majoron models we took into account, which provides spectral indices $n=$ 1, 2, 3, 7. The spectrum of the 2$\nu\beta\beta$ ($n= 5$) is also shown here. 
}
\label{theory}
\end{figure}

The total lepton number violation is a key element in the search for new 
physics evidences. Its observation would prove the existence of processes not described by the Standard Model and would give an important hint to explain the lack of antimatter in the known universe \cite{antimatter,baryonasym}. \\
The possibility that the neutrino is a Majorana particle brought a deeper interest into the neutrinoless double $\beta$ decay (\DBD) \cite{Furry_1939, Deppisch:2012nb, Doi1985, Primakoff_1959, Mohapatra1986, VERGADOS19861}, a promising process to search for the total lepton number violation. It is an alternative mode of the Standard Model allowed double $\beta$ decay (2$\nu\beta\beta$) \cite{GoeppertMayer} which occurs via the emission of 2 electrons and 2 neutrinos and presents a continuous energy spectrum ending at the Q-value (Q$_{\beta\beta}$). On contrary, the \DBD \ provides the emission of 2 electrons only which carry all the energy, resulting into a monochromatic peak at the Q$_{\beta\beta}$.
The $2\nu\beta\beta$ has been studied by several experiments and it is one of the rarest processes in the universe: its half life has been precisely measured and lies in the range 10$^{18}$--10$^{24}$ yr depending on the isotopes \cite{BARABASH2020}. On the other hand, the \DBD \ keeps eluding detection. At the present day, the experiments have a range of sensitivity spanning from 10$^{24}$ to 10$^{26}$ yr \cite{Agostini1445,PhysRevLett.117.082503,PhysRevC.100.025501,PhysRevLett.123.161802,PhysRevD.92.072011,PhysRevLett.123.032501,cuorenature,cupidmo2021,gerdacollaboration2020final}.\\
Other exotic modes of the double $\beta$ decay, which could be an evidence of new physics, have been conjectured over the years. In many of these cases, the energy spectrum of the 2 emitted electrons results to be a continuous function which is distorted with respect to the case of the 2$\nu\beta\beta$. Thus, it is possible to search for these modes through a detailed study of the energy spectrum shape \cite{lv,ssd,PhysRevD.93.072001,nemo3-dbd,PhysRevLett.122.192501,PhysRevD.98.092007,Armengaud2020_2nu}.\\
In particular, the assumption that the lepton number symmetry is spontaneously broken leads to the existence of a massless Goldstone boson, which in the original models was called the Majoron $\chi_0$ \cite{CHIKASHIGE1981265, GELMINI1981411,GEORGI1981297}.
Since then, the precision measurement of the width of the Z boson decay to invisible channels has strongly constrained the original models for the Majoron \cite{lep2006}, and in recent years many other models, free of this constraint, predicting light or massless Majoron-like particles have been proposed \cite{BAMERT199525,BEREZHIANI199299}. In these models the definition of Majoron is more general, refering to massless or light bosons that could be or not Goldstone bosons and differ by the leptonic number carried by the Majoron.
The Majoron(s) would be emitted in the final state of the neutrinoless double $\beta$ decay by coupling to the Majorana neutrino (\DBD $\chi_0$): 
\begin{linenomath}
\begin{equation}
    \begin{aligned}
    (A,Z) \rightarrow (A,Z+2) + 2e^- + \chi_0\\
    (A,Z) \rightarrow (A,Z+2) + 2e^- + 2\chi_0
    \end{aligned}   
\end{equation}
\end{linenomath}
The Majoron(s), escaping detection, would give rise to a missing energy, producing a characteristic distorted continuous spectrum. \\
The current Majoron models are divided in two classes depending on whether the decay violates the lepton number or it is balanced by the emission of a leptonically charged Majoron. The absence of the $0\nu\beta\beta$ would create a tension with the lepton number violating models. However, unless the $0\nu\beta\beta$ is observed, the leptonically neutral Majoron models can’t be distinguished from their counterparts in which the lepton number is conserved. Each model is characterized by a number of emitted Majorons, $m$,  and a different spectral index ($n=$ 1, 2 and 3 for m=1 and n = 3 and 7 for m=2), which determines the shape of the energy spectrum as follow:
\begin{linenomath}
\begin{equation}
    \frac{d\Gamma}{d\epsilon_1\,d\epsilon_2} \sim (Q_{\beta\beta}-\epsilon_1-\epsilon_2)^n
\end{equation}
\end{linenomath}
where $\epsilon_{1,2}$ are electron energies.
This is the most important characteristic from the experimental point of view since it allows to distinguish the \DBD $\chi_0$($\chi_0$) from the 2$\nu\beta\beta$, which has a spectral index $n=$ 5. 
The energy spectrum predicted by each of the considered models is shown in Fig. \ref{theory}.\\
The search for \DBD $\chi_0$ was performed in several nuclei, including $^{136}$Xe (EXO-200 \cite{exomajoron} and KamLAND-Zen \cite{Gando_2012}), $^{76}$Ge (GERDA \cite{gerdamajoron}), $^{100}$Mo (NEMO-3 \cite{nemo100mo}) and $^{130}$Te (CUORE \cite{cuoremajo}), resulting in half-life sensitivities ranging from 10$^{21}$ to 10$^{24}$ yr. The most recent search for this decay with the isotope $^{82}$Se was carried out by NEMO-3 \cite{Arnold_2018}, which reported an half-life lower limit of 3.7 $\times$ 10$^{22}$ yr in the case $n=$1. 
In this work, we present a new result for the search of \DBD $\chi_0$ in $^{82}$Se with the Phase I data of the CUPID-0 experiment.

\section{The CUPID-0 Experiment}
The CUPID-0 experiment collected data from 2017 to 2020 at the underground Laboratori Nazionali del Gran Sasso (LNGS, L'Aquila, Italy). It is the first demonstrator for CUPID (CUORE Upgrade with Particle IDentification), the next generation experiment for the search for \DBD \ with scintillating cryogenic calorimeters (also called bolometers) \cite{hallc2020,CUPID_preCDR_2019,optcupid}.
These detectors consist of crystals containing the isotope candidate for the decay. The working principle of cryogenic calorimeters relies on the detection of the temperature increase, i.e. heat release, due to an energy deposit in the crystal. This can be achieved thanks to the very low temperature reached in dedicated cryogenic facilities. The conversion of the temperature increase into an electric pulse happens through a cryogenic sensor, which is a Neutron Transmutation Doped germanium (NTD-Ge \cite{Hallerf}) thermistor. \\
Scintillating bolometers combines the excellent energy resolution ($<$1\% FWHM at few MeV of energy deposit) and containment efficiency (about 80-90\%) of cryogenic calorimeters with the particle identification capabilities offered by heat and scintillation light simultaneous read-out. CUPID-0 \cite{Azzolini_2018}, based on the experience achived by the LUCIFER \cite{Beeman:2012jd,Beeman:2012gg,Cardani:2013mja,Beeman:2013sba,Beeman:2013vda,Cardani_2013,Artusa:2016maw} and LUMINEU \cite{BEEMAN2012318,Barabash:2014una,Armengaud:2015hda,Bekker:2014tfa,Armengaud_2017,Grigorieva2017,Poda2017} projects, was the first experiment to explore the ultimate background suppression offered by this technique. \\
The detector consists of 26 cylindrical ZnSe scintillating crystals, 24 of which are enriched at 95\% level in $^{82}$Se, the candidate isotope for the \DBD. The scintillation photons escaping the crystals were detected by means of light detectors. These consist in high purity germanium disks working as thin cryogenic calorimeters. \\
Furthermore, the choice of the $^{82}$Se as candidate isotope for the \DBD \ was aimed at mitigating the $\gamma$ background since its Q-value (2997.9 $\pm$ 0.3 keV \cite{qse}) lies above the most significant peaks from natural radioactivity.\\
The CUPID-0 detectors are arranged in 5 towers, consisting in ZnSe crystals interleaved with light detectors: both are held by Polytetrafluoroethylene (PTFE) elements and thermally coupled to a copper structure. Each crystal is surrounded by a reflecting foil (VIKUITI{\texttrademark}) to enhance the light collection.\\

CUPID-0 was the first bolometric experiment to reach a background level of the order of 10$^{-3}$ counts\,/(keV\,kg\,yr) and it set the most stringent half-life limit of the $^{82}$Se \DBD \ to fundamental and excited states \cite{0vphase2cupid0, azzoliniprl,Azzolini_excited_states}. The sources contributing to the background in CUPID-0 data have been deeply analysed in Ref. \cite{Azzolini_2019}. This work provided a detailed knowledge about the 2$\nu\beta\beta$ spectral shape, making possible to prove the SSD (single state dominated) mode \cite{ssd} and to to search for CPT violation \cite{lv}. 

\section{Data Analysis}
\label{da}
\begin{figure}[htbp]
\centering
\includegraphics[scale=0.42]{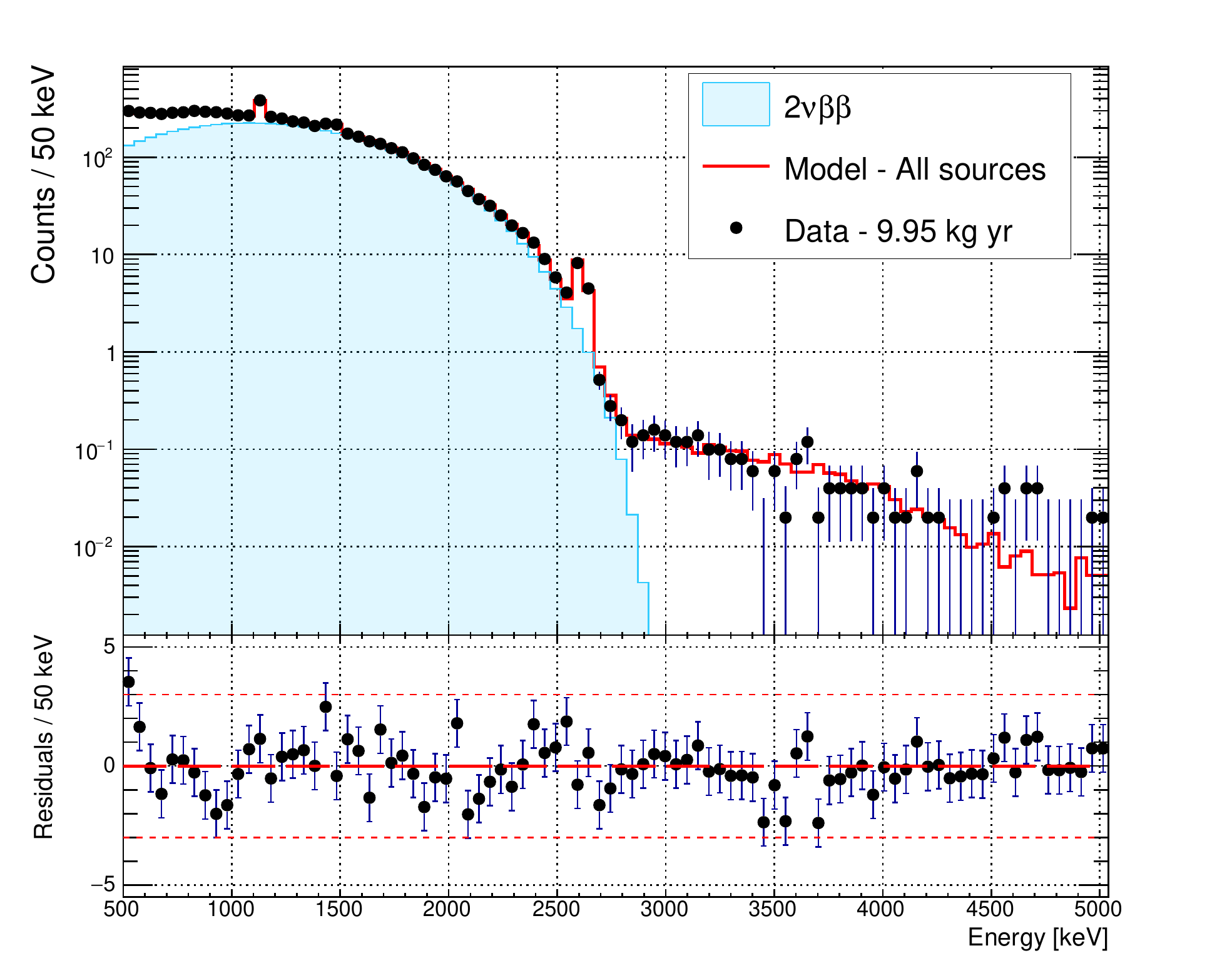}
\includegraphics[scale=0.42]{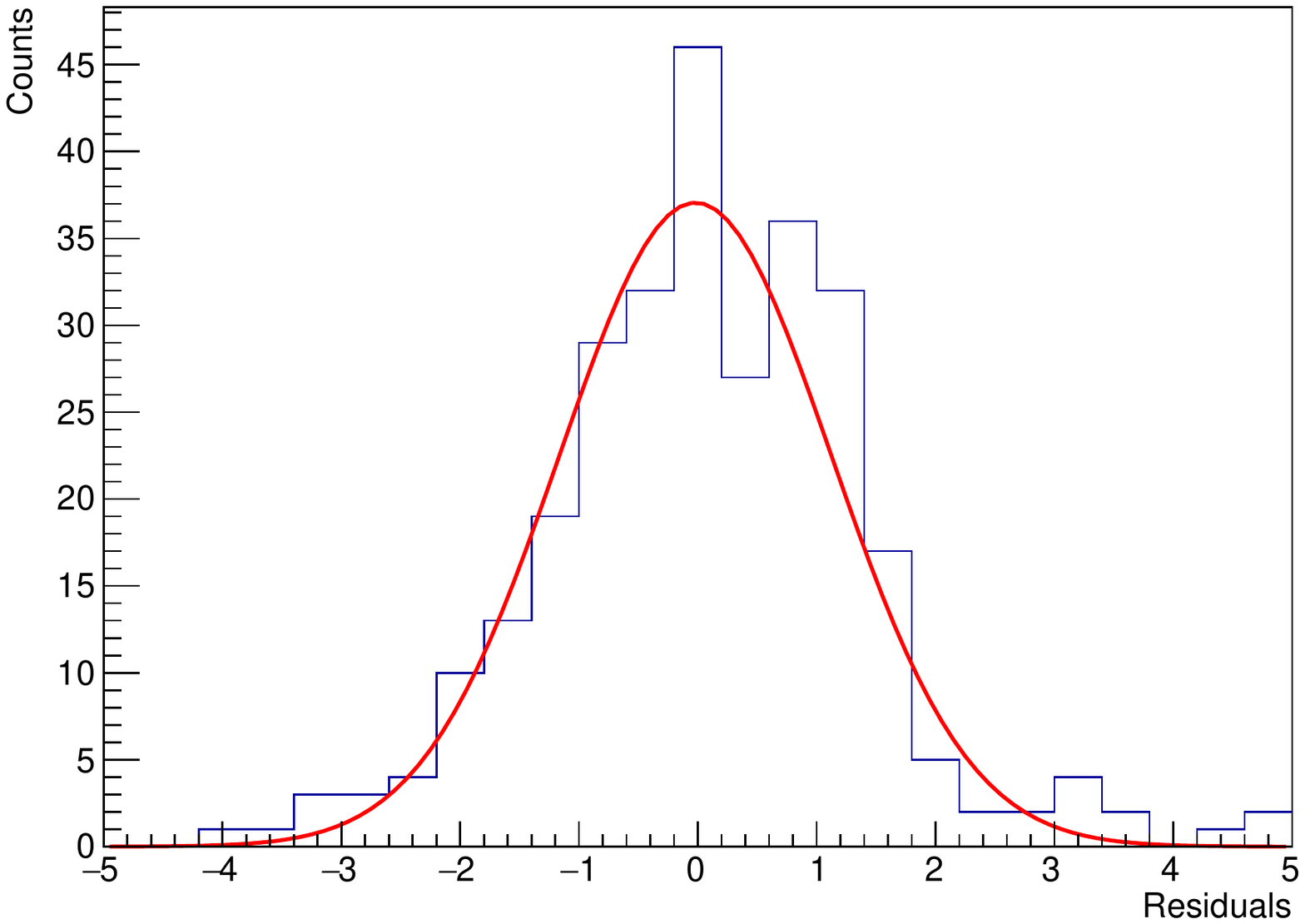}
\caption{In the top panel we show the $\mathcal{M}_{1\beta/\gamma}$ data spectrum and fit result (Model - All Sources) in background only hypothesis with the residual corresponding to each bin. For illustrative purposes, we show here the spectrum with 50\,keV fixed steps bins, but the the fit results have been obtained with a variable size binning as outlined in \ref{da}. The dominant background component, the $^{82}$Se 2$\nu\beta\beta$, is superimposed in blue. In the bottom panel we show the distribution of the residuals arising from the fit and the best fit Gaussian function in red. Its mean resulted to be -0.02 $\pm$ 0.07 and the width 1.15 $\pm$ 0.06.}
\label{pull}
\end{figure}

The set of data analysed in this work represent the CUPID-0 Phase I (running from June 2017 to December 2018), in which we collected a total Zn$^{82}$Se exposure of 9.95\,kg $\times$ yr (i.e. 3.41 $\times$ 10$^{25}$ $^{82}$Se nuclei and 8.74\,kg of $^{82}$Se active mass).\\
The detailed data processing and selection is described in \cite{Azzolini:analysis:2018}. As a quantity of interest for this work, we only report here that the selections to exclude non-particles events led to a total efficiency of $\epsilon=$ (95.7 $\pm$ 0.5)\,\%, constant above 150\,keV \cite{Azzolini_2019}.\\
As a starting point for this analysis we needed to reconstruct the background sources contribution to the CUPID-0 energy spectrum. This study was performed in Ref. \cite{Azzolini_2019} by using the JAGS software. For this work we reproduced these results, obtaining consistent results for all the sources, by means of a different software, i.e. BAT (Bayesisan Analysis Toolkit, \href{http://mpp.mpg.de/bat/}{bat.mpp.mpg.de}), which we then used to implement the algorithm for the \DBD$\chi_0$ search. 
We evaluated the list of sources starting from the $\alpha$ and $\gamma$ lines in the data spectrum \cite{Azzolini_2019}. These are mainly due to contamination in the volume of the ZnSe crystals and on their surfaces, namely $^{65}$Zn, $^{40}$K, $^{60}$Co, $^{147}$Sm, $^{238}$U/$^{232}$Th decay chains and the $^{82}$Se 2$\nu\beta\beta$ decay. For the surfaces contamination, we considered two different levels of depth inside the crystals (10\,nm and 10 $\mu$m) and we adopted the same strategy for the reflecting foils contamination (from $^{238}$U/$^{232}$Th decay chains). Then, we introduced other potential sources which do not produce prominent signatures in the spectrum. These are due to $^{238}$U/$^{232}$Th and $^{60}$Co contamination inside and outside the cryostat and in the Roman lead shield, which is placed inside the cryostat itself. We took into account the effect of the presence of these contaminants in our results as a systematic error (see Sec. \ref{n1}). Finally, to account for the environmental background, we included the cosmic muons as a source, while the $\gamma$-rays and neutrons contributions are expected to be negligible \cite{Alduino_2017}. The energy spectrum of the sources was simulated with a Monte Carlo (MC) method.\\
The model we used to describe the data consists in a set of 33 scale parameters corresponding to each background source. The scale parameters work as weights to the MC spectra. In particular we normalized the MC spectrum so that the value of the scale parameter corresponds to the number of events produced by a given source. It follows that the activity of each source in the CUPID-0 array is directly related to the corresponding scale parameter through the livetime, the number of emitting nuclei and the detector efficiency.\\
We adopted a Bayesian approach to evaluate the sources activities which best fit to data. The likelihood of the model consists of a Poisson distribution of the number of events in each bin of the energy spectrum, combining together all the scale parameters. \\
Then, we set a flat and non-negative prior probability for each scale parameter, with few exceptions. In particular, we set a gaussian prior, centered in 0 and non-negative, on the $^{232}$Th source present in the reflecting foils, as its decay product ($^{228}$Ra) contamination was constrained by independent measurements (while we didn't include $^{238}$U because it was discarded) \cite{Azzolini_2019}. Then we set gaussian priors to the $^{60}$Co of the cryostat, as measured in the same facility by CUORE-0 \cite{Alduino_2017}, and on the cosmic muons, which are constrained from data exploiting the signature given by events coincident in multiple-crystals. Moreover, we set a gaussian prior on the bulk-to-surface events ratio of $^{226}$Ra-$^{210}$Pb decay chain from $^{238}$U and $^{228}$Ra-$^{208}$Pb from $^{232}$Th by constraining these from parent-daughter nuclei time correlated events \cite{delayedcupid0}.\\
In order to better constrain the sources activities, we split the CUPID-0 data into 4 sets, each used to build an energy spectrum. First, we identified coincident events triggered in multiple crystals of the array within a time window of 20\,ms. The choice of the time window was optimized by identifying events from the 2615\,keV line (from a $^{232}$Th source) which trigger two ZnSe crystals. In this way, we obtained $\mathcal{M}_1$ and $\mathcal{M}_2$ spectra for single and double hit events respectively.\\ 
Then, exploiting the particle identification capabilities of the detector \cite{Azzolini:analysis:2018}, we divided $\mathcal{M}_1$ by identifying events due to $\beta/\gamma$ ($\mathcal{M}_{1\beta/\gamma}$ see Fig. \ref{pull}) and $\alpha$ particles ($\mathcal{M}_{1\alpha}$). 
The $\beta/\gamma$--$\alpha$ discrimination was performed only above 2\,MeV, where the light detectors performance is good enough to ensure an efficient ($>$99.9\%) particle identification \cite{Azzolini_2019, Azzolini_2018}. Below 2\,MeV the unidentified $\alpha$ particles were added to the $\beta/\gamma$ spectrum. \\
Finally, we considered for the fit the spectrum $\Sigma_2$, in which we summed the energy of the coincident events. We applied the same procedure of data selection for both data and MC simulations. We performed a simultaneous fit by summing the log-likelihoods defined for each of the 4 spectra. \\
We sampled the joint posterior probability with a Markov Chain Monte Carlo method exploiting the Metropolis algorithm implemented in the BAT software.\\
 \\
We performed the fit with a low energy threshold of 500\,keV and a variable step size binning in order to contain any peak present in the data spectrum in a single bin \cite{Azzolini_2019}. This choice helps in mitigating the effect of the detector response on the model. \\
As a figure of merit for the result of the fit, we show in Fig. \ref{pull} the distribution of the residuals for each bin of the 4 fitted spectra. We evaluated the uncertainty for each bin combining a Poissonian contribution with the properly propagated scale parameters uncertainties from the fit result. If data and the model are in agreement, we expect that the residuals are distributed as a normalized Gaussian. We found the best fit Gaussian function to be compatible with a normal, with a mean of -0.02 $\pm$ 0.07 and a width of 1.15 $\pm$ 0.06, thus confirming the good data--model agreement resulting from the fit. 

\subsection{Results on \DBD$\chi_0$ with n = 1}
\label{n1}
\begin{figure}[htbp]
\centering
\includegraphics[scale=0.45]{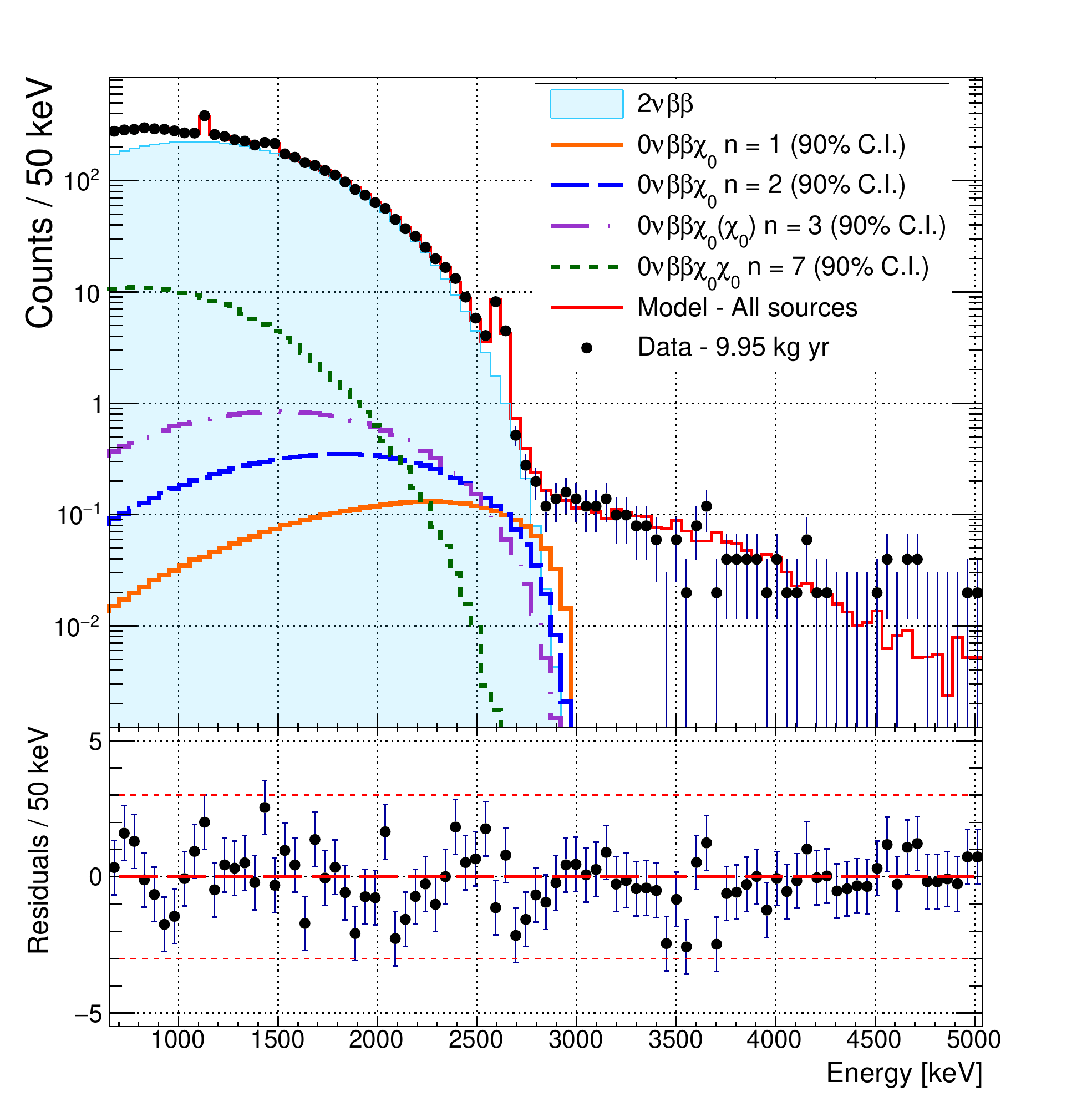}
\caption{In this plot we report the data and the best fit model spectra in the case of \DBD$\chi_0$ with $n=$ 1. The \DBD$\chi_0$ with $n=$ 1 spectra at 90\% C.I. limit and the 2$\nu\beta\beta$ are superimposed. For illustrative purposes we also show the 90\% C.I. limit of the spectra for the \DBD$\chi_0$($\chi_0$) with $n=$ 2, 3 and 7. In the bottom panel we show the difference between the number of events in each bin of the data and of the best fit model normalized to the bin uncertainty (in the reference fit context with the $n=$ 1 \DBD$\chi_0$). }
\label{datamodel}
\end{figure}

The background only model can be easily extended to include an extra source due to an exotic process such as the Majoron emission.  As for the 2$\nu\beta\beta$, the signature for this decay is given by 2 electrons depositing energy in a single crystal, producing a continuous energy spectrum. However, the expected spectral shape is modified, and this feature makes possible to reconstruct the number of events due to \DBD $\chi_0$ and to 2$\nu\beta\beta$. In this section we will focus on the search for Majoron models presenting a spectral index $n=$ 1, describing the adopted analysis strategy. \\
We added a free parameter, with a flat and non-negative prior, to the background model previously described to account for the \DBD $\chi_0$ together with the background sources.
We performed the fit considering an energy threshold of 700\,keV for the $\mathcal{M}_{1\beta/\gamma}$ spectrum. This choice removes the possible contribution of most of the pure $\beta$ emitter contamination, which is correlated with the signal and highly anticorrelated with the 2$\nu\beta\beta$. Nevertheless, we studied how the low energy threshold affects the result of this search. We performed the fit moving the threshold to 500, 300 and 200 keV and we found that the sensitivity on the \DBD$\chi_0$ with n = 1 is reduced by lowering the threshold (see ``Threshold" in Table \ref{Table:sys}). \\
As in the background only hypothesis, we choose a variable step size binning. We also performed the fit using a fixed step binning of 15, 30 and 50\,keV (see ``Binning" in Table \ref{Table:sys}). As the binning width decreases, the agreement with data worsen, as the fit can't reproduce the peaks spectral shape well enough. \\
In the following, we will refer to the choices of a 700\,keV threshold and a variable step binning, together with the full list of 33 background sources as the ``reference" model. The corresponding spectrum is shown in Fig. \ref{datamodel}

\begin{figure}[htbp]
\centering
\includegraphics[scale=0.45]{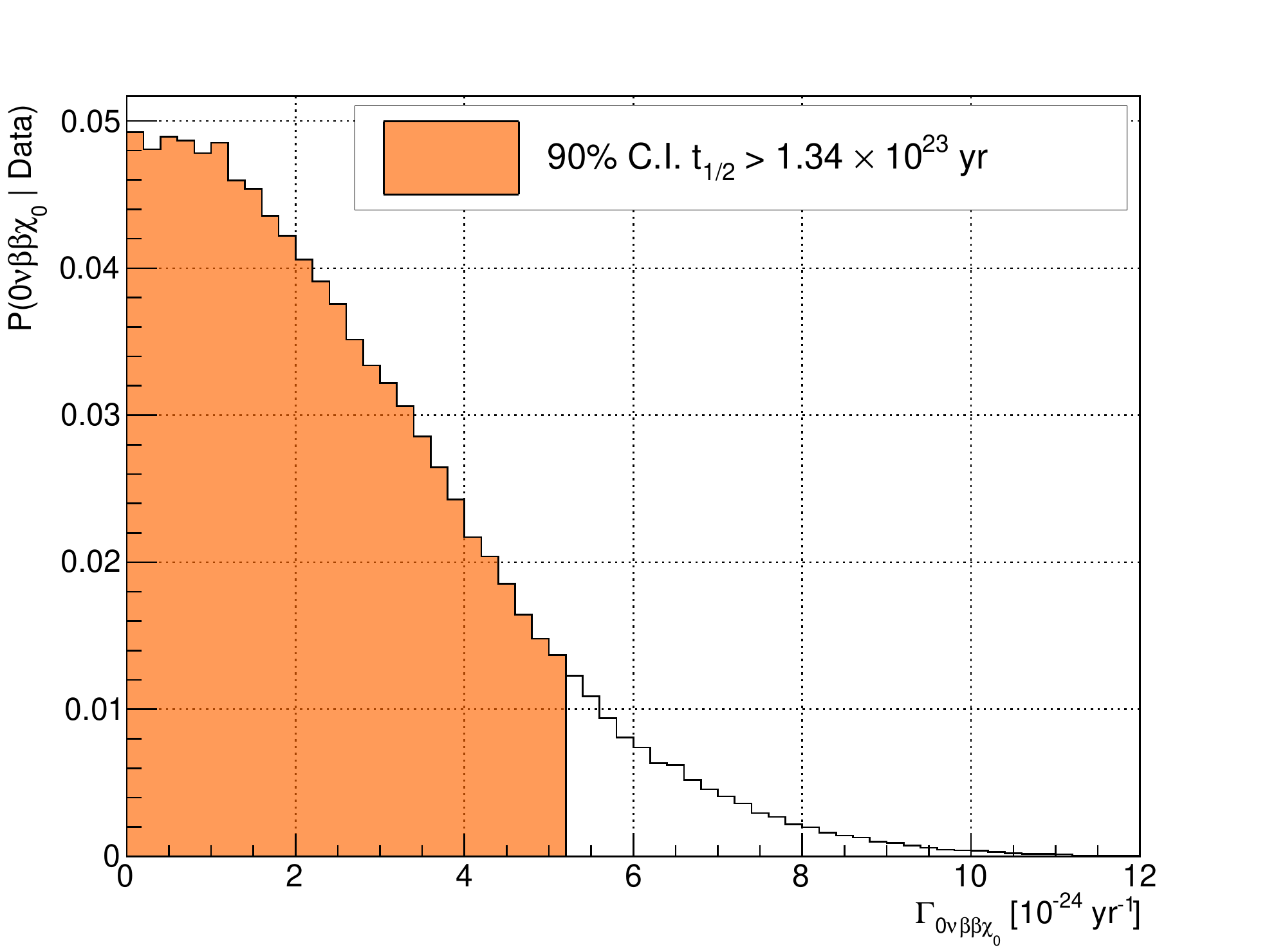}
\caption{Posterior of the rate of the \DBD$\chi_0$ with $n=$1 spectral index in the case of the reference fit. The orange area corresponds to the 90\% of the total integral of the posterior.}
\label{posterior}
\end{figure}

\begin{figure}[htbp]
\centering
\includegraphics[scale=0.42]{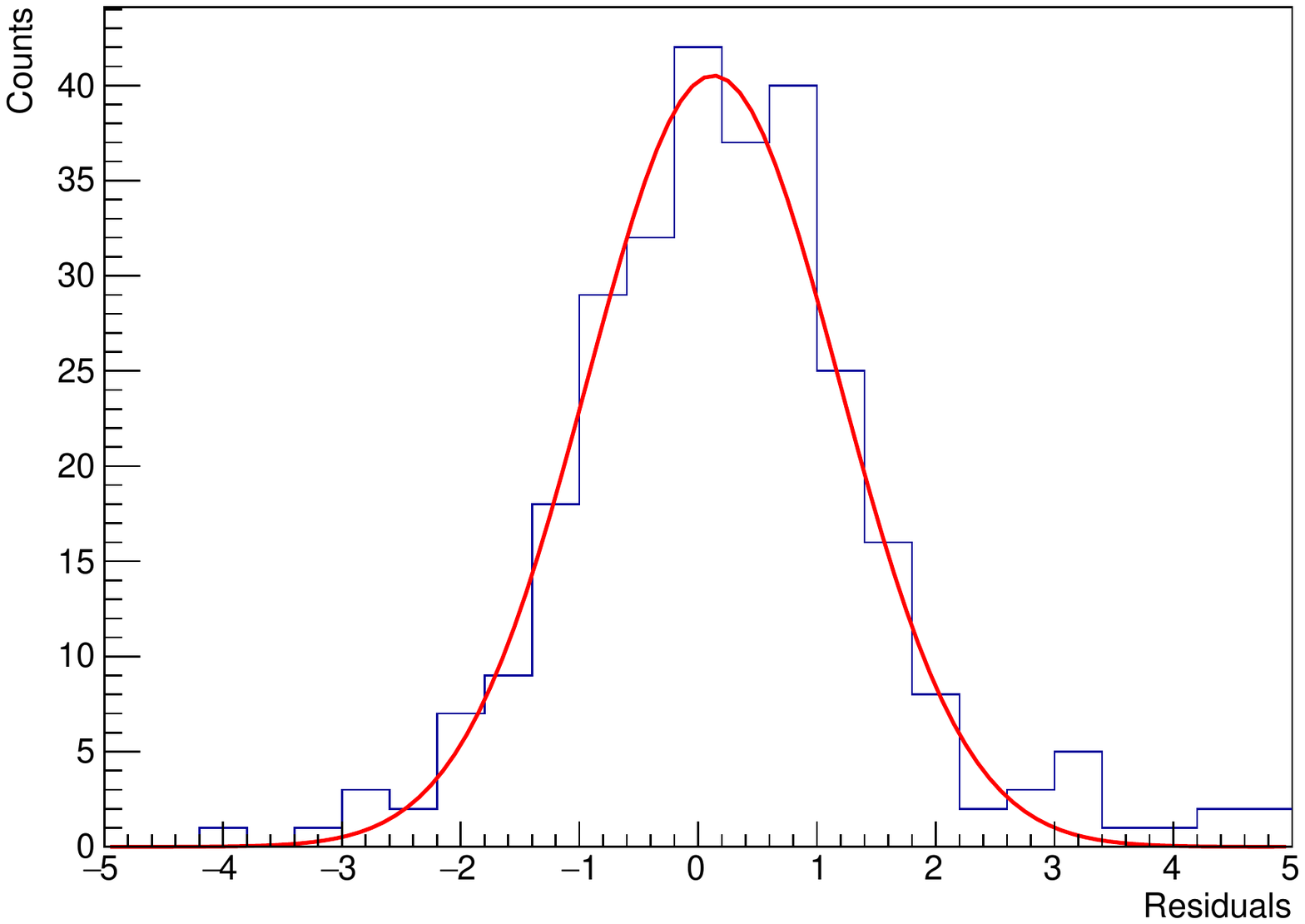}
\caption{Residuals distribution in the ``reference" model with the $n=$ 1 \DBD$\chi_0$. The best fit Gaussian function is shown in red. Its mean resulted to be 0.13 $\pm$ 0.07 and the width 1.06 $\pm$ 0.05.}
\label{pullM1}
\end{figure}

The number of events assigned by the reference model fit to the \DBD$\chi_0$ spectrum is 90 $\pm$ 68. So the activity of the \DBD$\chi_0$ is compatible with 0 within 1.3$\sigma$. 
Thus we set a lower limit at 90\% of credibility interval (C.I.) on the \DBD$\chi_0$ half-life of 1.34 $\times$ 10$^{23}$ yr (Fig. \ref{posterior}). We verified that the good data--model agreement is confirmed also including the \DBD $\chi_0$ as a source. Indeed, by fitting the residuals distribution, we found a mean of 0.13 $\pm$ 0.07 and  a width of 1.06 $\pm$ 0.05, as shown in Fig. \ref{pullM1}.\\
However, although the data--model agreement quality is unaffected, the introduction of a new spectrum in the fit could influence the activity estimation of other sources. This effect is expected to be larger for the most correlated (or anticorrelated) sources. In this case, due to its spectral shape, the 2$\nu\beta\beta$ source presents the highest anti-correlation with the \DBD$\chi_0$. Nevertheless, the estimation of the 2$\nu\beta\beta$ activity is compatible within 1.1 $\sigma$ with the result obtained in the background-only hypothesis (reported here \cite{Azzolini_2019}). \\
\\
To study the dependence of the result on systematic errors effects we repeated the fit procedure changing the model several times. One of the most important systematic effects in this study is the potential presence of a pure $\beta$ emitter as a contaminant source. Thus, we included in the fit a scale parameter corresponding to a $^{90}$Sr contamination source, which is produced by nuclear fission. This isotope decays to the $^{90}$Y which is itself a $\beta$ emitter with a Q-value higher than 2\,MeV. This is the only pure $\beta$ emitter contamination producing events above the threshold we used in the fit. The $^{90}$Sr-$^{90}$Y contamination produces a continuous spectral shape which anticorrelates with the 2$\nu\beta\beta$. On the other hand, the $n=$1 \DBD $\chi_0$ is correlated with the $^{90}$Sr-$^{90}$Y (and anticorrelated with the 2$\nu\beta\beta$). Thus, the introduction of a pure $\beta$ emitter in the fit reduces the half-life sensitivity on the signal process, as we found from the fit (see ``$^{90}$Sr-$^{90}$Y" in Table \ref{Table:sys}).\\
We took into account as a systematic error the sources location in the cryogenic facility (see ``Source Location" in Table \ref{Table:sys}). To test this systematic effect, we removed from the fit the $^{60}$Co and the $^{40}$K sources inside (``No Internal") or outside (``No External") the cryostat. Similarly, we also removed the $^{232}$Th/$^{238}$U inside, outside the cryostat or in the Roman Pb shield (``No in Roman Pb"). We then performed the fit removing all the sources which were converging at 0 in the reference fit (see ``Less Sources" in Table \ref{Table:sys}). We concluded that the fit is robust enough under reasonable changes in the list of sources taken into account, as the variation on the signal half-life sensitivity is lower than 16\% in all the cases we tested.\\
Finally we studied the systematic error related to the energy calibration (see ``Energy Scale" in Table \ref{Table:sys}). We applied a different calibration choice to data which takes into account the residuals computed in \cite{PhysRevLett.123.032501}. By shifting the energy scale to higher values, we expect that the fit will attribute more events to the \DBD$\chi_0$ since its spectral shape is peaked close to the endpoint (the Q$_{\beta\beta}$). This would reduce the half-life sensitivity while, on the contrary, would be enhanced by shifting the scale to lower energies. We verified this effect by shifting the energy scale by a constant value corresponding to the maximum (about +3\,keV) and minimum (about --5\,keV) residual. Then we corrected the energy scale by applying the whole residuals curve, which is a parabolic function. The overall effect is a reduction of the sensitivity of about 16\%.\\

\begin{table}[htbp]
\centering
\caption{Limit at 90\% C.I. on the half-life of the \DBD$\chi_0$ with $n=$ 1. The different results correspond to the tests we performed to understand the effect of the systematic errors. The result achieved in the Reference model configuration is also reported here.}
\begin{tabular}{lccc}

\hline\hline
Model && t$_{1/2}$ 90\% C.I (10$^{23}$ yr) \\
\hline
\hline
Reference &&  \textgreater1.34  \\
\hline
$^{90}$Sr-$^{90}$Y &&   \textgreater1.00   \\
\hline
Energy Scale &&  \textgreater1.13   \\ 
\hline
Source Location &   &\\
No External&    &\\
 &$^{60}$Co &   \textgreater1.29   \\
 &$^{40}$K &   \textgreater1.34  \\
 &$^{232}$Th/$^{238}$U &   \textgreater1.13   \\
No Internal&   &\\
 &$^{60}$Co &  \textgreater1.45    \\
 &$^{40}$K &   \textgreater1.31   \\
 &$^{232}$Th/$^{238}$U &   \textgreater1.31   \\
No in Roman Pb&    &\\
&$^{232}$Th/$^{238}$U &   \textgreater1.34   \\
\hline
Less sources &&   \textgreater1.26  \\
\hline
Binning & &\\
&15\,keV &   \textgreater2.11   \\
&30\,keV &   \textgreater1.92  \\
&50\,keV &   \textgreater2.40  \\
\hline
Threshold & &\\
&500\,keV &   \textgreater1.25   \\
&300\,keV &   \textgreater1.07  \\
&200\,keV &   \textgreater1.01\\
\hline
\hline
\end{tabular}
\label{Table:sys}
\end{table}

\subsection{Results on \DBD$\chi_0$($\chi_0$) with n = 2,3 and 7}
Once we finalized the fit strategy and we studied the systematic effects, we searched for the \DBD \ with the emission of Majoron-like particles predicted by other possible models. All these models lead to only three possible spectral indices different from 1, i.e. $n=$ 2, 3, 7 (Fig. \ref{theory}). As a remark, the case with $n=$ 7 exists only with the emission of 2 Majorons (\DBD$\chi_0\chi_0$), $n=$ 3 can happen with 1 or 2 Majorons while $n=$ 1 or 2 with a single Majoron only. So we replaced one by one the $n=$ 1 \DBD$\chi_0$ spectrum in the fit sources list with the $n=$ 2, 3, 7 spectra. \\
In all the cases we verified that the data--model agreement is satisfactory and that the \DBD$\chi_0$ (or \DBD$\chi_0\chi_0$) activity is compatible with 0 within at most 1.2$\sigma$. So again we set a limit at the 90\% of  of the corresponding posterior. The half-life lower limits we found in the reference fit context resulted to be 4.8 $\times$ 10$^{22}$, 2.1 $\times$ 10$^{22}$ and 2.1 $\times$ 10$^{21}$ yr for $n=$ 2, 3 and 7 respectively.\\
The half-life limits are less stringent as the spectral index increases since the spectral shape gets more similar to the 2$\nu\beta\beta$ one. Thus, the anti-correlation between the \DBD$\chi_0$ (or \DBD$\chi_0\chi_0$) and the 2$\nu\beta\beta$ increases with the spectral index, leading to a broader posterior, and a reduction of the Majoron search sensitivity.\\ 
We also verified that the estimate of the 2$\nu\beta\beta$ activity is compatible with the value reported in \cite{Azzolini_2019} in all the cases. In particular this estimate is more affected with the introduction of the \DBD$\chi_0\chi_0$ with $n=$ 7 because of its high anti-correlation factor (about --0.9). In the latter case, the 2$\nu\beta\beta$ activity resulted to be (9.86 $\pm$ 0.08) $\times$ 10$^{-4}$ Bq/kg, which is compatible within 1.2 $\sigma$ with the estimate in the background only hypothesis.\\ 
As done for the $n=$ 1 spectrum, we repeated the fit changing the model to study possible systematic effects. This study, if applied on spectra with indices $n=$ 2 and 3, brought us to similar conclusions as in the case $n=$ 1. Instead, the \DBD$\chi_0\chi_0$ with $n=$ 7 produced different results since, unlike the $n=$ 1,2 and 3 cases, the peak of the \DBD$\chi_0$ spectrum lies at lower energies with respect to the 2$\nu\beta\beta$. This induces a different behaviour under some changes in the model or in the fit parameters. For instance, moving the energy threshold to lower values improves the limit since it enhances the peak structure, reducing in this way the anti-correlation with the 2$\nu\beta\beta$. Moreover, by introducing the $^{90}$Sr-$^{90}$Y pure $\beta$ emitter source, we observed a more stringent limit, contrarily to what observed for $n=$ 1, 2 and 3. Indeed, while the other Majoron spectra are correlated with the $^{90}$Sr-$^{90}$Y, the $n=$ 7 one is anti-correlated with it, because its spectral shape is peaked at lower energies. This changes the correlation relationships in the fit, causing a lower number of events assigned to the \DBD$\chi_0\chi_0$, and thus a more stringent limit. \\
The result of the search of the \DBD$\chi_0$($\chi_0$) with indices $n=$ 1, 2, 3 and 7 is summarized in Fig. \ref{datamodel}, where the spectra at the 90\% C.I limit are reported together with the data and the best fit model for the case $n=$ 1 in the context of the reference model. The bottom panel shows the difference between the number of events of the data and the reference model for the case $n=$1 normalized to the bin uncertainty, i.e. the residuals. 

\section{Discussion}
To take into account the systematic effect on the search for the \DBD$\chi_0$, we decided to combine all the posteriors obtained in the different tests we performed (see Table \ref{Table:sys}) by following the law of total probability. So we considered the reference fit just as one of the possible choice of the model to describe the data behaviour. So we summed all the posterior weighting them by the prior probability of the model. We assigned an equal prior to each of the families (Reference, $^{90}$Sr-$^{90}$Y, Energy Scale, Source Location, Less sources, Binning and Threshold) of tests listed in Table \ref{Table:sys}, as done in Ref. \cite{lv}. The results are reported in Table \ref{Table:cc}. We also tried different approaches to weight the posteriors: we assigned an equal prior to each test, regardless of the family, and we evaluated the prior of each model based on the data--model agreement. However, these approaches led to minor changes (on the order of 1\%) in the final results, so we opted for the approach already proposed in Ref. \cite{lv}.\\
\begin{table}[htbp]
\centering
\caption{Results in terms of half-life and Majoron-neutrino effective coupling constant, achieved for the \DBD$\chi_0$ search with each of the Majoron models considered. }
\begin{tabular}{cccc}
\hline\hline
Decay & n & \ \ t$_{1/2}$ 90\%C.I. (yr) \ \ & $\lvert \left \langle g_{\chi_{0}} \right \rangle \rvert $\\
\hline
\DBD$\chi_0$ & 1 &  \textgreater1.2 $\times$ 10$^{23}$ & \textless (1.8--4.4)$\times$10$^{-5}$ \\
\DBD$\chi_0$ & 2 &  \textgreater3.8 $\times$ 10$^{22}$ & -- \\
\DBD$\chi_0$ & 3 &  \textgreater1.4 $\times$ 10$^{22}$ &   \textless0.020   \\
\DBD$\chi_0\chi_0$ & 3 &  \textgreater1.4 $\times$ 10$^{22}$ &  \textless1.2       \\
\DBD$\chi_0\chi_0$ & 7 &   \textgreater2.2 $\times$ 10$^{21}$ &   \textless1.1\\
\hline
\hline
\end{tabular}
\label{Table:cc}
\end{table}\\
The lower limit obtained for the different Majoron-emitting double $\beta$ decays can be converted into an upper limit on the  Majoron-neutrino coupling constant using:
\begin{linenomath}
\begin{equation}
    [ t_{1/2} ]^{-1} = |\left \langle g_{\chi_{0}} \right \rangle |^{2m} \, G_{(m,n)}^{(0)} \, g_A^4 \, |M_{(m,n)}|^2
\end{equation}
\end{linenomath}
where $t_{1/2}$ is the \DBD$\chi_0$ (or \DBD$\chi_0\chi_0$) half-life, $g_{\chi_{0}}$ is the Majoron-neutrino coupling constant, $m$ is the number of Majorons in the final state, $G_{(m,n)}^{(0)}$ the phase space factor, $g_A$ is the weak axial coupling constant and $M_{(m,n)}$ the nuclear matrix element (NME). As above, n is the spectral index of the decay mode. By computing $g_{\chi_{0}}$ it is possible to directly compare the results achieved with different candidate isotopes. 
\begin{figure}[htbp]
\centering
\includegraphics[scale=0.33]{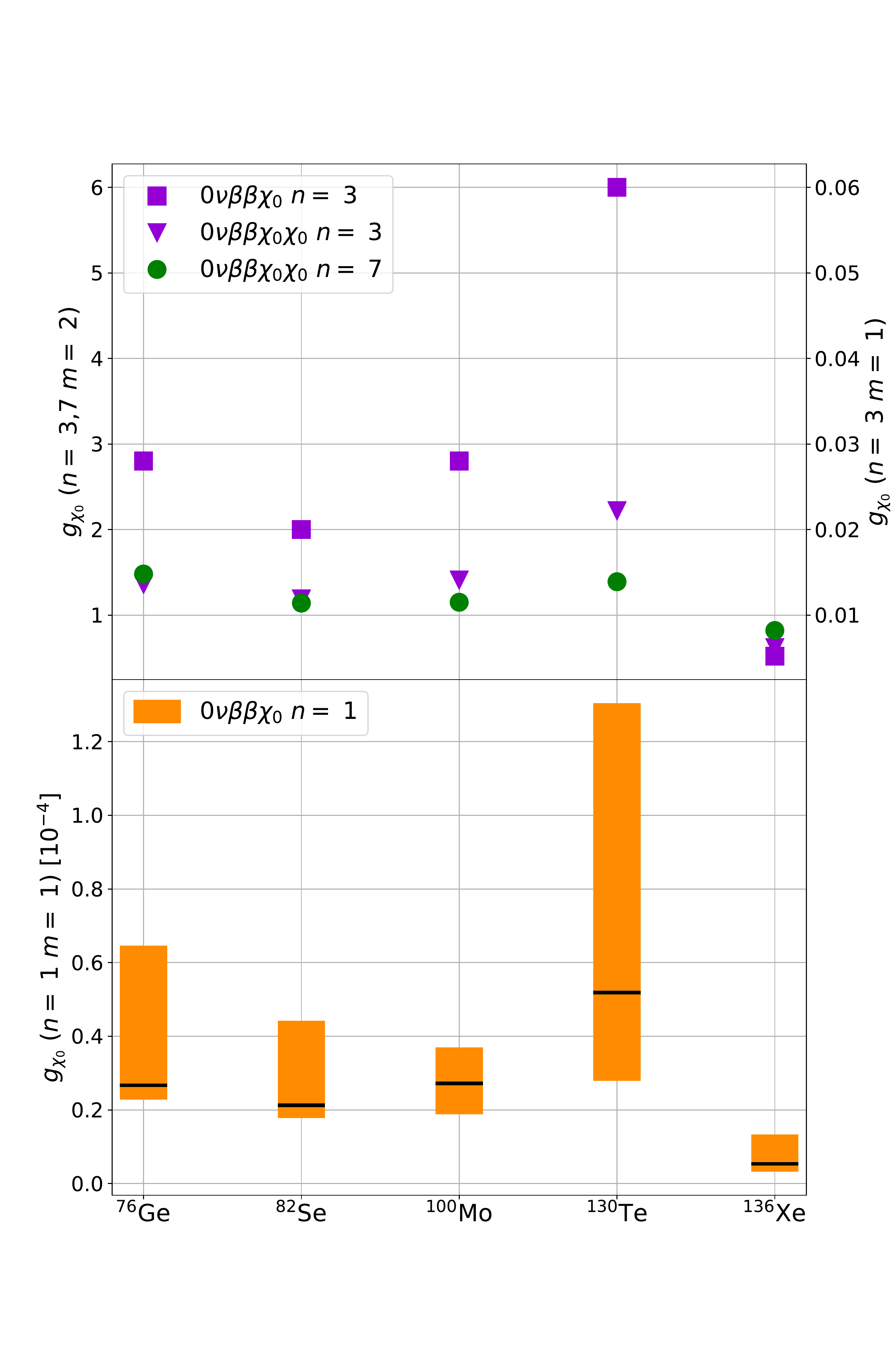}
\caption{Present best limits on the Majoron--neutino effective coupling constant for each isotope. In the bottom panel we show the limits for the \DBD$\chi_0$ with $n=$ 1, the error bands are due to NMEs range. The black lines represent the results achieved with the NMEs reported in \cite{Kotila_2021}, from which we took also the NMEs to compute the $g_{\chi_{0}}$ upper limit for the decay modes with $n=$ 3 and 7 and $m=$ 1 or 2 shown in the top panel. The best limits values come from GERDA \cite{gerdamajoron} for the $^{76}$Ge (with an exposure of  20.3\,kg $\times$ yr), from this work for the $^{82}$Se (exposure 9.95\,kg $\times$ yr), from NEMO-3 \cite{nemo100mo} for the $^{100}$Mo (exposure 34.3\,kg $\times$ yr), from CUORE \cite{cuoremajo} for $^{130}$Te (exposure 216.4\,kg $\times$ yr) and from EXO-200 \cite{exomajoron} for the $^{136}$Xe (exposure 234.1\,kg $\times$ yr).  }
\label{limits}
\end{figure}\\
The NMEs for the decay mode with $n=$ 1, which coincide with the ones of the \DBD, have been computed in several theoretical frameworks as the Nuclear Shell Model \cite{Menendez_2009,Men_ndez_2017,Horoi2016,Coraggio_2020}, Interacting Boson Model \cite{JK2015,Deppish2020,Kotila_2021}, Quasi-particle Random Phase Approximation \cite{Mustonen2010, Simkovic2013, Simkovic2018, Hyvarinen2015,Fang2018,Terasaki_2020}, Energy-Density Functional theory \cite{Rodriguez2010,lopezvaquero2013, Song2017} and other methods \cite{Rath2013, Yao2015}. 
All these computations result into a wide range of possible NMEs, which is reflected into the estimate of $g_{\chi_{0}}$, as represented by the yellow bands in the bottom panel of Fig. \ref{limits}. We considered the bare value of $g_A$ reported in each NME reference. The black lines shown in the plot correspond to the values of $g_{\chi_{0}}$ computed with the NMEs from \cite{Kotila_2021}, where are also reported the NMEs we used in the case of the decay modes with $n=$ 3 and 7 and $m=$ 1 or 2 (shown in the top panel of Fig. \ref{limits}). The phase space factor we used are reported in \cite{Kotila_2015}. Instead, in the case of $n=$ 2, there are no NME theoretical computations available. As a reference, we also report in this plot the upper limit on the Majoron-neutrino coupling constant achieved by other experiments which investigated different candidate isotopes. The limits we show here are the lowest achieved so far for each of the isotopes considered. These comes from the half-life limits reported in Ref. \cite{gerdamajoron,nemo100mo, cuoremajo, exomajoron} and in this work in the case of $^{82}$Se, combined with the phase space factors and NMEs ranges described above for each isotope. The $g_{\chi_{0}}$ limits we found with the data of CUPID-0 Phase-I are reported in Table \ref{Table:cc}. \\
In conclusion, despite the relatively low exposure, we found a competitive result which is coherent with the search for Majoron-like particles performed with different isotopes. 
\\

\section*{Acknowledgments}
This work was partially supported by the European Research Council (FP7/2007-2013) under Low--background Underground Cryogenic Installation For Elusive Rates Contract No. 247115. We are particularly grateful to M. Iannone for the help in all the stages of the detector construction, A. Pelosi for the construction of the assembly line, M. Guetti for the assistance in the cryogenic operations, R. Gaigher for the calibration system mechanics, M. Lindozzi for the development of cryostat monitoring system, M. Perego for his invaluable
help, the mechanical workshop of LNGS (E. Tatananni,
A. Rotilio, A. Corsi, and B. Romualdi) for the continuous help in the overall setup design. We acknowledge the Dark Side Collaboration for the use of the low-radon clean room. This work makes use of the DIANA data analysis and APOLLO data acquisition software which has been developed by the CUORICINO, CUORE, LUCIFER, and CUPID-0 Collaborations

\bibliography{biblio.bib}

\end{document}